\documentclass[12pt]{article}

\begin{document}\setlength{\unitlength}{1mm}

\begin{titlepage}

\begin{flushright}
{RUNHETC-2011-02}\\
\end{flushright}

\vskip2cm
\begin{center}
{\large {\bf A Note On Measuring Charm and Bottom}}
\\
{\ }
\\
{\large {\bf Forward-Backward Asymmetries at the Tevatron}}
\vskip1.5cm {Matthew~J.~Strassler}
\vskip1cm
{\it New High Energy Theory Center}\\
{\it Department of Physics and Astronomy,\\
Rutgers, 136 Frelinghuysen Rd, Piscataway, NJ 08854, USA\\
}

\end{center}

\vspace{0.5cm}
\begin{abstract}
The forward-backward asymmetry $A_{FB}^{t}$ in top quark production at
the Tevatron has been seen to be anomalously large both by CDF and D0.
Parton-level asymmetries as large as 50\%, with a large error bar,
have been extracted from the data.  It is important to measure other
quark asymmetries if possible, as these would help clarify the source
of any new physics behind $A_{FB}^t$.  In this note it is argued that
asymmetries in $b\bar b$ and $c\bar c$ should be accessible to the
Tevatron experiments, using the full data sets.  A crude study
suggests that muon asymmetries in high-$p_T$ dijet events, with
suitable use of muon and jet kinematics and (inefficient) heavy flavor
tagging, might allow detection of $A_{FB}^c,A_{FB}^b\sim$ 0.3.  Were
it possible to make heavy flavor tagging at high $p_T$ efficient, or
mistags rare, then the sensitivity of the measurement of $A_{FB}^b$
could be significantly better.
\end{abstract}

\end{titlepage}

Recent measurements by both the CDF and DZero experiments at the
Tevatron \cite{CDFafb,D0afb,CDFnew} have shown a persistent and
anomalously large forward-backward asymmetry in top quark production.
This exciting discrepancy with the Standard Model has generated a
substantial amount of theoretical activity 
explaining the effect.

The asymmetry is observed mainly in events in which the $t\bar t$
invariant mass is above 450 GeV \cite{CDFnew}.  The observed top quark
asymmetries have been converted into a $t\bar t$-frame partonic-level
asymmetry, the quantity that can be directly extracted from a
Lagrangian at leading order.  This inferred asymmetry in \cite{CDFnew}
is very large, though with a substantial error bar: $A_{FB}^t=0.475\pm
0.114$, far above the next-to-leading-order QCD expectation.

If this anomalous asymmetry is real, and is due to new physics
affecting the third generation more generally, a similar effect might
impact the bottom quark.  In particular one would expect this if new
physics affects the $t_L$, and perhaps even if it only affects $t_R$.
Or perhaps the asymmetry lies in the up-quark sector and affects the
charm quark.  It is also possible that only the $t_R$, of all quarks,
is affected, and no anomalous asymmetry is present in either bottom or
charm quarks.  Thus, the measurement of bottom and charm quark
asymmetries is important in diagnosing any new physics that may be
generating $A_{FB}^t$.  These observables are also attractive in that
a different set of issues are involved, compared to the $A_{FB}^t$
case, in converting a lab-frame or parton-frame measured asymmetry to
the intrinsic partonic asymmetry, the essential quantity for comparing
with theory.\footnote{The extraction of $A_{FB}^t$ in the partonic
  frame is highly non-trivial, as it requires understanding angular
  distributions of the final state particles, various detector
  effects, and the process by which the $t\bar t$ events are
  reconstructed.  It seems very difficult, without precisely repeating
  the CDF analysis, to appreciate fully the subtleties involved.}

The purpose of this note is to argue that the charm and bottom
forward-backward asymmetries, $A_{FB}^c$ and $A_{FB}^b$, might be
experimentally accessible at the Tevatron, using methods similar
to those suggested
long ago in \cite{oldpaper}, with the full data sets at
D0 and CDF.\footnote{While this study was underway, a paper appeared
  \cite{newpaper} that also suggests that the measurement of
  $A_{FB}^b$ is both important and possible.}  We will check that a 10
inverse fb data set gives each experiment a sample of dijet events
with a single non-isolated muon that is large enough to make such
measurements plausible.

To determine the combination of techniques required for these
difficult measurements lies far beyond the reach of a non-expert; all
that will be attempted here is to motivate the measurement by showing
the possibility of success.  In particular, only a very cursory study
is performed, with many limitations, and one could certainly do a
better job, accounting properly for jet reconstruction, missing energy
from neutrinos, next-to-leading-order effects, and a tracing of
systematic errors.  However, to be more precise and accurate would
require a detailed understanding of the multivariate heavy-flavor
taggers that the experimentalists have designed, among other
subtleties.

Indeed a very large uncertainty, especially for $A_{FB}^b$, comes from
the unknowns in heavy-flavor tagging.  The most experimentally
interesting region, for comparison with what is observed in
$A_{FB}^t$, is at high invariant mass.  But tagging at high $p_T$ (of
order 150--250 GeV) poses many challenges \cite{neutag}, so far little
studied, as statistics at these $p_T$s is low, and few measurements at
these energies have required heavy flavor tagging.\footnote{The author
  thanks Y.~Gershtein and G.~Watts for discussions on this point.} The
results presented below therefore have very large error bars from this
source.

The essential experimental ingredients in measuring these asymmetries
are the kinematics of the jets and of the muons embedded within them,
heavy flavor tagging using displaced vertices and tracks, and the
measurement of the charges of the muons, which are correlated with the
charges of the produced quarks.\footnote{Electrons will not be
  considered here, as they are difficult to detect efficiently in jets
  at CDF and D0.  Their inclusion might help marginally, but perhaps
  this issue is best left to the experimentalists.}  The method for
measuring an asymmetry boils down to the following:
\begin{itemize}
\item Select dijet events, at high invariant mass $m_{jj}$ and low to moderate
$|\eta|$, that contain at least one embedded muon.
\item Use  variables such as $p_T^{jet}$, $p_T^{rel}$ (the transverse momentum of the
muon relative to the jet) and $z_\mu=p_T^\mu/p_T^{jet}$ (the fractional
momentum of the jet carried by the muon) to separate, statistically,
the different sources of muons.
\item Use track-based heavy-flavor tagging information within the two jets to 
change the mix of sources for the observed muon.
\end{itemize}
Then observe the forward-backward asymmetry of the $\mu^-$  events (combined
with the backward-forward asymmetry of the $\mu^+$
events) in the dijet frame.

Let us take a moment to recall all the reasons why the muon observed
in a jet does not directly correlate to its parent parton's charge.  
Charm quarks
generally produce a muon with the same charge as the parent quark.
But bottom quarks do not, because (a) mixing of neutral $B$ mesons
converts $b\bar q\to \bar b q$, and (b) a cascade decay in which a
bottom hadron decays hadronically to a charm hadron, which then decays
semileptonically, produces a muon of opposite charge to the $b$
charge.  A
gluon can split to $b\bar b$ and $c\bar c$, so a muon from an ensuing
heavy-flavor decay has a random charge.\footnote{The same is largely true for
muons that come from decays of light hadrons in flight, though this
will be irrelevant because of their small numbers in event samples
with rather hard muons.  Fake muons from punchthrough have a charge
bias, but this should be forward-backward symmetric.}

To explore whether the asymmetry measurements are feasible, let us
step through the results of a crude but instructive study using Pythia
\cite{PYTHIA}.  The numbers below will not be accurate for many
reasons (K factors are not included, jets are not actually
reconstructed, effects of the neutrinos are not accounted for, many
experimental subtleties are not considered, Pythia is not entirely
trustworthy in heavy flavor production and decay, etc.) but are
intended to be illustrative and motivate more careful experimental
studies.

Let us start with a pure dijet sample, with a cut on $m_{jj}$ of 450
GeV (above which scale
$A_{FB}^t$ is observed to be anomalously large) and requiring the
$p_T$ of the leading jet be above 150 GeV (which is high enough to
fire a dijet trigger.)  The leading-order cross-section is about
300 pb.  Within this sample, select the events with at least one
non-isolated muon of $p_T>20$ GeV and $|\eta|<2$.  The corresponding
cross-section for this preselection sample is of order 1.8 pb.
At CDF, where the muon system
extends only out to $|\eta|\sim 1.1$, an acceptance factor of order
$0.85$ should be included in the following discussion.

Note that the muon is not needed for triggering.  An alternative
strategy at D0 would be to also include muon-triggered events out to
$|\eta|=1.6$, with a lower requirement on the $p_T$ of the jet.  This
might marginally increase statistics, though it runs the risk that, if
the asymmetry decreases too fast at low $m_{jj}$, the signal itself may
decrease to outweigh any statistical advantage.  The optimal method
will depend on triggering considerations and will require detailed
study.

Some effort will be required to ensure muons from top and electroweak
processes are negligible or are carefully removed.  Top quark pair
production can largely be rejected by event shape and jet counting,
while electroweak processes such as $W$ plus jets and single top,
important backgrounds as they are significantly forward-background
asymmetric, must be modeled, using the kinematic regions where the
muon from the $W$ is isolated and extrapolating to where it is not.
We will assume this can be done with sufficient confidence.  An
irreducible asymmetric background comes from $d\bar d \to s\bar
c,\ \bar s c$ via $W$ exchange.  This appears to be too small to
impact the low-precision measurements of asymmetries considered here,
but should be removed when converting the measurement to a limit or
observation of $A_{FB}$.

The preselected 1.8 pb sample is divided roughly into\footnote{The
  fragmentation functions for $b$ and $c$ hadrons within $b$ and $c$
  jets are not too well known at high $p_T$, and Pythia perhaps makes
  the distribution of the muon $p_T$ a bit too hard, which would cause
  an overestimate of the signal.  However, conversely, the
  preselection criteria chosen here are not optimized, and the muon
  $p_T$ cut used here might be higher than necessary. This will
  require detailed study using data.}
\begin{itemize}
\item 1.0 pb of $qq$, $q\bar q$, $qg$, $\bar qg$, and $gg$ scattering, of which 0.25 pb 
creates a $gg$ final state,
\item 0.35 pb of $qQ$, $\bar qQ$, $q\bar Q$, $\bar q\bar Q$, $gQ$, $g\bar Q$ (where $Q=c,b$)
scattering, of which 0.05 pb involves a gluon in the final state,
\item 0.15 pb of $c\bar c$ production, of which about 5\% is from
$gg$ initial states,
\item 0.3 pb of $b\bar b$ production, of which about 5\% is from 
$gg$ initial states.
\end{itemize}
The muons in these events come predominantly from heavy flavor if a
heavy quark is present, with a rate just above $10\%$ per $b$ quark,
and about half of this for charm.  (About a third of the muons in a
$b$ jet are ``wrong-sign'', opposite in sign to the $b$ quark charge.)
In the other events, they come mostly from gluons splitting to bottom
and charm quarks.  About $2.5\%$ of the muons come from other sources
(such as decays of lighter hadrons in flight.)  

A number of events
have two muons; most of these are in the same jet.  Some come from
gluon splitting to two heavy mesons.  About half the di-muon events come from $b\bar b$,
and less than half of these have one muon in each jet.  Because the
various di-muon subsamples are all small, their importance to the
measurement is marginal, and in any case too complex to be 
investigated here.  We set them aside for a more careful analysis to
consider properly.

Now let us turn to the asymmetries.
Somewhat remarkably, we are best off first considering $A_{FB}^c$.  With
10 inverse fb, the preselected sample contains about 18000 events, including
about 1500 $c\bar c$ events.  Since $\sqrt{18000}\sim 134$,
there is already 3$\sigma$ statistical sensitivity to an $A_{FB}^c\sim 0.3$.

The situation for $A_{FB}^b$ is slightly worse.
The dilution of the true asymmetry by wrong-sign
muons in $b$ jets reduces the observed asymmetry by a factor of
about three.  Without any additional effort, there is statistical sensitivity only
to $A_{FB}^b\sim 0.45$.

Given that the top quark asymmetries are claimed to be perhaps as
large as 50\%, these simple measurements are already of considerable
scientific interest, and they don't even require use and understanding
of heavy-flavor tagging at high $p_T$.  But the drawback of this very
simple approach is that the actual observed asymmetries are very
small: statistical significance of $3\sigma$ corresponds
to an asymmetry of order $2\%$.  It
is easy to imagine that any observation would be complicated by
concerns about systematic errors.

We should therefore ask the following questions.  
First, can we enhance the purity of the $c\bar c$ or
$b\bar b$ sample, or decrease the wrong-sign muon contribution in $b$
jets, so as to increase the statistical sensitivity of the
measurement?  If not, can we at least maintain the sensitivity while
increasing the observed size of the asymmetry, potentially reducing
systematic effects?  And can we find a control sample that will allow
increased confidence that an observed asymmetry is due to the
underlying particle physics?

With regard to the first two questions, a number of different attempts and
considerations suggest that improved sensitivity in $A_{FB}^b$ (but
not $A_{FB}^c$) is possible, while a larger observed asymmetry with
comparable sensitivity is possible in both cases.  A sketch of these
arguments now follows.

First, let us consider the obvious method of heavy-flavor tagging,
whose usefulness is very sensitive to its poorly known effeciency and
fake rate.  For scale, imagine that heavy-flavor tagging were as good
as it is at lower $p_T$.  Suppose we applied tagging to the other jet
(the jet with no muon, or perhaps the jet with the lower $p_T$ muon if
both jets contain one).  If we were to assume, naively, a $50\%$
flavor tagging efficiency for bottom quarks, a $15\%$ rate for charm
quarks, a $1\%$ mistag rate for light quarks and a 3\% mistag rate for
gluons (remembering that the gluon has a substantial probability to
split to bottom quarks and charm quarks, and quoted mistag rates have
already backed this probability out) we get a sample that is $75\%$
pure $b\bar b$, with a total cross-section of 0.2 pb.  This improves
the sensitivity for $A_{FB}^b$ by a factor of more than 2, immediately
putting a 3$\sigma$ measurement of $A_{FB}^b\sim 0.2$ within reach,
with an observed $A_{FB}^b$ (accounting for dilution both by
backgrounds and wrong-sign muons) of about 5\%.  Meanwhile, the
sensitivity to a charm asymmetry significantly decreases.

Unfortunately, heavy-flavor tagging is not likely to work nearly so well at
high $p_T$.  A more conservative estimate,
such as a $30\%$ $b$-tagging rate and a $5\%$ mistag rate, would almost
eliminate the gain in efficiency.  The reality at CDF seems to lie
somewhere between \cite{neutag}.

Tagging both jets simultaneously is not helpful.  The cost in signal
is too high.\footnote{In making estimates, it is important to keep in
  mind that the flavor mix of the jet containing the muon is different
  from that of the other jet.}  But tagging {\it at least one} jet is
useful.  Even with the conservative tagging just mentioned, it appears
a gain in sensitivity of 1.5 would be possible.  This alone would
bring sensitivity to $A_{FB}^b$ up to the initial sensitivity to $A_{FB}^c$.

There are alternatives to tagging.  Measurement of $A_{FB}^b$ (but not
$A_{FB}^c$) can be improved by reducing the number of wrong-sign
muons.  There are two natural avenues: demand a large $p_T^{rel}$,
or demand a large $z_\mu = p_T^\mu/p_T^{jet}$.  The former is more
commonly used, but when both jets and muons have large $p_T$, the
measurement of a small $p_T^{rel} \sim 2$ GeV requires knowing the jet
angle precisely.  The resolution on this quantity may become a
problem\footnote{The author thanks G.~Watts for discussions on this
  point.} at these energies, especially in the presence of the
neutrino that accompanies the jet.  The variable $z_\mu$ (after the
missing $p_T$ from the neutrino is added back into the jet $p_T$) may
not suffer as badly at high $p_T$.  Demanding a cut on $z_\mu$ of
order 0.2 or 0.25 seems to improve sensitivity to $A_{FB}^b$ by about
20\% while increasing the size of the observed asymmetry by about a
factor of 2.

Another method to enhance the asymmetry (but not the statistical
sensitivity) is to raise the cut on the jet $p_T$.  (One could also
consider a cut on the muon $p_T$, but this is obviously correlated
with $z_\mu$ and jet $p_T$, which are themselves largely
uncorrelated.)  This cut reduces the backgrounds from processes with
initial-state gluons and/or heavy-quarks.  As an example, the 
current study suggests that a jet $p_T$ cut of 225 GeV reduces the signal
by about 1/2 and background by about a 1/4.  Sensitivity is very
slightly degraded, for both $A_{FB}^b$ and $A_{FB}^c$, but any
observed asymmetry would nearly double in size.

Finally, in measurements of this type it is important to have control
samples, in which no new-physics asymmetry is expected.  One could
lower the muon $p_T$ cut below 20 GeV, but experimental study is
required to understand what new backgrounds would enter into that
sample.  Within the original preselection sample, we have already
found that raising jet $p_T$ and $z_\mu$ cuts ought to increase any
partonic asymmetries, and so therefore reversing the jet $p_T$ and
$z_\mu$ cuts would provide control samples.  Looking at low $z_\mu$
enhances wrong-sign muons and should reduce any $A_{FB}^b$ to an
unmeasurable degree.  Looking in the lower jet $p_T$ bins reduces both
$c\bar c$ and $b\bar b$ relative to backgrounds, and any asymmetries
should disappear.  These control samples consist of about half the
original pre-selection set, with somewhat less than 10000 events.
Obtaining higher-statistics control samples will be challenging.

Taking these ideas together brings us to the following tentative
conclusions.  A preselection along the lines of what was suggested
here allows already for some sensitivity.  Suppose an asymmetry is
observed; then the hypothesis that it comes from heavy flavor requires
that it come dominantly from the higher jet-$p_T$ bins (and the
central region in $\eta$.)  If it does, then the question is whether
it is from $b$ or $c$.  Methods to reduce wrong-sign muons from $b\to
c\to\mu$ cascades, such as the use of $z_\mu$ or $p_T^{rel}$ of the
muon, enhance $A_{FB}^b$ but hurt $A_{FB}^c$.  The usefulness of
track- and vertex-based tagging within the two jets --- in particular
the requirement that at least one of the two jets be so-tagged ---
depends very sensitively on the details of tagging and mistagging at
high $p_T$.  If tagging works well, it can improve the sensitivity to
$A_{FB}^b$ by a factor of more than 2; if it works poorly, it might
still give improved sensitivity of perhaps a factor of 1.5. Either
way, tagging can help determine whether an asymmetry comes from an
underlying $A_{FB}^b$ or from an underlying $A_{FB}^c$, since standard
tagging will reduce $A_{FB}^c$ to unobservable levels.\footnote{A
  specialized tagging strategy that could enhance charm at such high
  $p_T$ while sufficiently rejecting both bottom and other sources of
  muons seems difficult to imagine, but is clearly worth some
  additional thought.}  How to optimize these methods will require
much more detailed study than is possible here.  But it seems likely
that, relative to the preselected sample's 3$\sigma$ sensitivity of
$A_{FB}^c\sim 0.3$ and $A_{FB}^b\sim 0.45$, which corresponds to an
observed asymmetry of order 2\%, the observed asymmetries can be
further enhanced by a factor of at least 2 without a loss of
sensitivity; and for $A_{FB}^b$, the sensitivity can probably be
improved by a factor between 1.5 and 2.5.

It is natural to ask whether measurements of this type are possible at
the LHC.  This issue requires a separate study, 
but seems very challenging, for many reasons.  The sources
of $b\bar b$ and $c\bar c$ are largely from $gg$, and have no
asymmetry in the parton frame.  Even for those events that stem from
$q\bar q$ initial states, the $pp$ hadronic initial state requires one
to measure the direction of the $q$ in the initial state
statistically, using the boost of the partonic frame.  There are huge
$gg\to gg$ backgrounds that produce muons from gluon splitting, and
large boosted sources of $qQ$ and $gQ$ scattering.  
Furthermore, the fact that boosted events are required, and that the
initial $pp$ state carries a definite charge, could interplay with
detector angular acceptance in tricky ways that would be difficult to
untangle.  The advantage of the Tevatron for this particular
fqmeasurement is its relative clarity and simplicity. Perhaps some
progress may be made at LHC by requiring a muon in each of the two
jets, and using the huge statistical advantage that the LHC will
eventually possess at these energy scales.

It has been argued here that measurements at the Tevatron of
forward-backward asymmetries of order 30\% in charm and/or bottom
production should be detectable at both experiments, with the
possibility of further improvement for $A_{FB}^b$ if one is optimistic
regarding tagging at high $p_T$.  The results of this article are
crude, with large and unquantifiable error bars, and need to be
reconsidered carefully by the Tevatron experiments.  Even if the
conclusions of this article prove qualitatively correct, these
measurements will be difficult.  However, the scientific benefits to
carrying out these measurements --- null tests of the Standard Model,
and indeed of a very wide class of theories beyond the Standard Model
--- would make them worthwhile to perform.

\

The author thanks J.~Boudreau, N.~Craig, Y.~Gershtein, G.~Salam,
S.~Schnetzer, J.~Thaler, S.~Thomas, and G.~Watts for comments and
conversations.  The author thanks M.I.T. and
Boston University for hospitality.
This work was supported by NSF grant PHY-0904069
and by DOE grant DE-FG02-96ER40959.


\begin{thebibliography}{99}

\bibitem{CDFafb}
  T.~Aaltonen {\it et al.}  [CDF Collaboration],
   ``Forward-Backward Asymmetry in Top Quark Production in  $p\bar{p}$
  Collisions at $\sqrt{s}$ =1.96 TeV,''
  Phys.\ Rev.\ Lett.\  {\bf 101}, 202001 (2008)
  [arXiv:0806.2472 [hep-ex]]


\bibitem{D0afb}
  V.~M.~Abazov {\it et al.}  [D0 Collaboration],
   ``First measurement of the forward-backward charge asymmetry in top quark
  pair production,''
  Phys.\ Rev.\ Lett.\  {\bf 100}, 142002 (2008)
  [arXiv:0712.0851 [hep-ex]].

\bibitem{CDFnew}
  T.~Aaltonen {\it et al.}  [The CDF Collaboration],
  ``Evidence for a Mass Dependent Forward-Backward Asymmetry in Top Quark Pair
  Production,''
  arXiv:1101.0034 [hep-ex].

\bibitem{oldpaper}
  L.~M.~Sehgal and M.~Wanninger,
  ``Forward-Backward Asymmetry in Two-Jet Events:  Signature 
of Axigluons in Proton--Anti-proton Collisions,''
  Phys.\ Lett.\  B {\bf 200}, 211 (1988).

\bibitem{newpaper}
  Y.~Bai, J.~L.~Hewett, J.~Kaplan and T.~G.~Rizzo,
  ``LHC Predictions from a Tevatron Anomaly in the Top Quark Forward-Backward
  Asymmetry,''
  arXiv:1101.5203 [hep-ph].

\bibitem{neutag}
  C.~Neu  [CDF Collaboration],
  ``CDF b-tagging: Measuring efficiency and false positive rate,''
See also http://www-cdf.fnal.gov/physics/new/top/2004/btag/ .



\bibitem{PYTHIA}
  T.~Sjostrand, S.~Mrenna and P.~Z.~Skands,
  ``PYTHIA 6.4 Physics and Manual,''
  JHEP {\bf 0605}, 026 (2006)
  [arXiv:hep-ph/0603175].

\end{thebibliography}
\end{document}